\documentclass{aa}
\usepackage{epsfig}
\usepackage{natbib}
\bibpunct{(}{)}{;}{a}{}{,}

\begin{document}
\title{The Torino Observatory Parallax Program: 
	White Dwarf Candidates}
\author{ R.~L.~Smart, M.~G.~Lattanzi, B.~Bucciarelli, G.~Massone,
R.~Casalegno, G.~Chiumiento, R.~Drimmel, L.~Lanteri, F.~Marocco \& A.~Spagna
} 

\institute{Istituto Nazionale di Astrofisica (INAF), Osservatorio
Astronomico di Torino, Strada Osservatorio 20, I-10025 Pino Torinese,
Italy}

\offprints{Richard~Smart, \email{smart@to.astro.it}}
\date{Accepted 17-03-2003. Received 21-01-2003}

\authorrunning{Smart et al.}
\titlerunning{TOPP White Dwarf Candidates}

\abstract { We present parallax determinations for six white dwarf
candidates in the Torino Observatory Parallax Program. The absolute
parallaxes are found with precisions at the 2-3 milliarcsecond level.
For WD 1126+185 we find a distance incompatible with being a white
dwarf, implying an incorrect classification. For WD 2216+484 we find
our distance is consistent with a simple DA white dwarf rather than a
composite system as previously proposed in the literature. In general
it is found that the published photometric distance is an overestimate
of the distance found here.

}

\maketitle

\section{Introduction}
With the introduction of precise and efficient detectors such as the
CCD and the Multichannel Astrometric Photometer \citep{GAT87} the
determination of parallaxes from the ground have undergone a
revolution.  There are now many programs that routinely find relative
parallaxes with precisions of a few milliarcseconds (mas) see
\citealt{DAH98} for a review.  The largest of these programs is the
USNO Flagstaff program (\citealt{MON92A}, hereafter MON92) that uses a
dedicated 1.5 reflector with a number of CCDs and routinely produces
parallaxes with 1 mas precision. We have recently completed the Torino
Observatory Parallax Program (TOPP){ ~using an on site telescope}
specifically targeted at faint stars ($m_v > 11$) using the USNO
program as a model. This program will complement the HIPPARCOS and
USNO projects, producing comparable precision but for stars in the
intervening magnitude range.

In this article we discuss the instrumentation, observational
procedures, calibrations, reduction techniques and first results for
six white dwarf candidates of the TOPP list.

\section{Instrument and Observation Procedures}
The 1.05m REOSC telescope of the Osservatorio Astronomico di Torino
(OATo) is a long focus (994.2 cm) large plate scale (20.7$''$/mm)
reflecting telescope with an astrometrically corrected field of view
of 45$'$. The telescope is a scaled-down replica of the USNO 1.5m
telescope with a flat secondary mirror and a parabolic primary mirror.
Observations are made using a large format 1296x1152 pixel EEV
CCD05-30 with a per pixel scale of 0$''$.47 and a field of view of
10$'$x9$'$.  All observations are carried out through a standard Cousins I
filter; this was done to reduce the effects of refraction but has the
benefit of also being the zone of maximum quantum efficiency of the
telescope/CCD system.

\begin{center}
\begin{figure*}[htp]
\epsffile{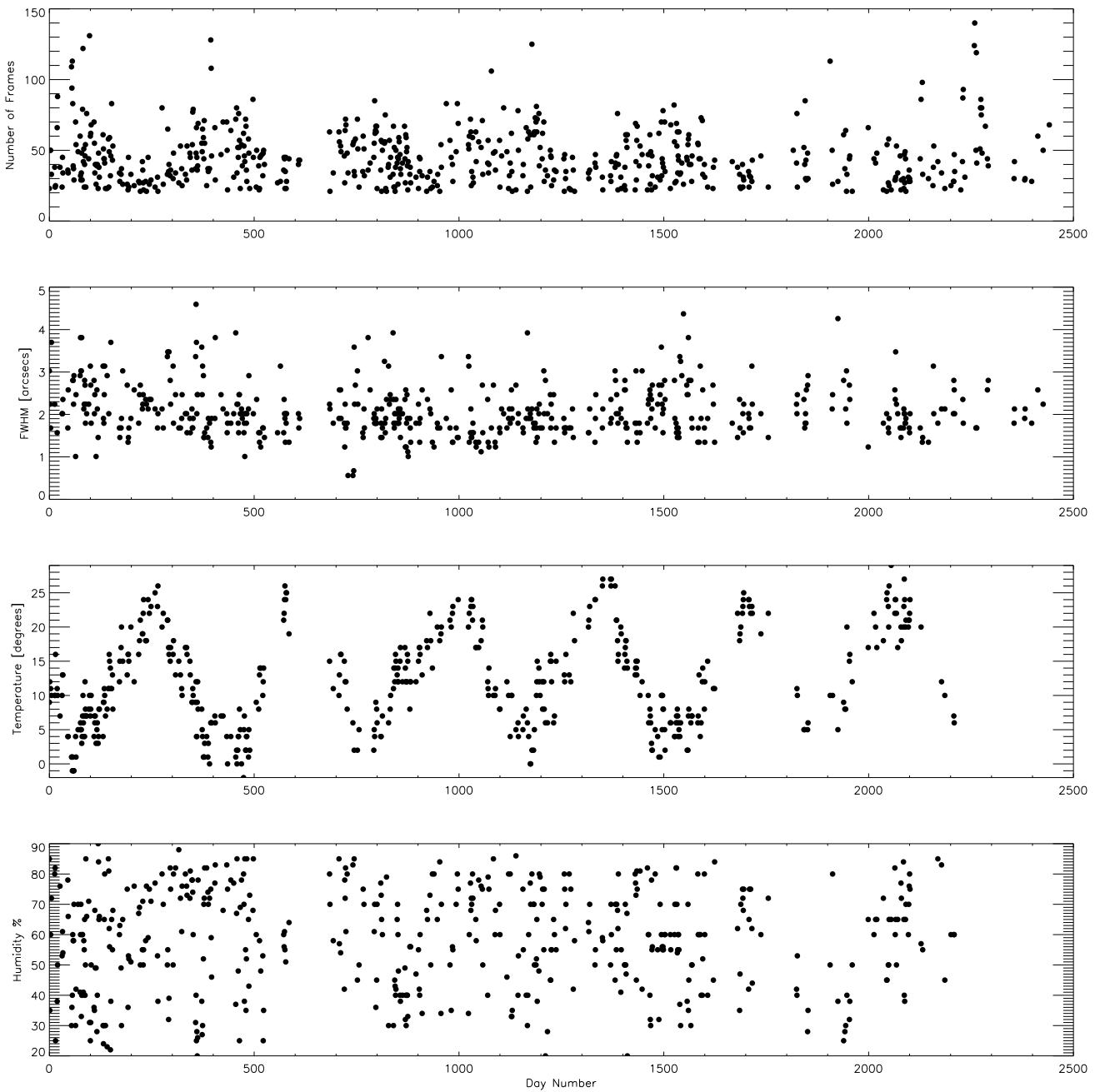}
\caption{The nightly number of frames taken, seeing, humidity and
temperature conditions. Day number is from the 11 November 1994, i.e. 
Julian date 2449667.0.}
\label{leggi_summary_1.ps}
\end{figure*} 
\end{center}

This telescope, filter and CCD combination was fixed, giving us a
stable system which is required for astrometric studies of this
nature.  The TOPP was provided with 40\% of the
observational time on this telescope during the period 10 November
1994 to 2 July 2001. In Fig. \ref{leggi_summary_1.ps} we plot the
change in temperature, humidity, seeing and number of frames observed
over the course of this period. In July 2001 the CCD was changed and
the program for these targets was halted. The telescope is
presently undergoing automation refitment and when
that is stable we will begin another program with the new, higher
resolution, CCD.

During the observational period the only major maintenance was a
periodic realuminisation of the mirrors on 14/04/1997, 24/05/1999 and
04/12/2000.  This required a dismounting of the mirrors and
transportation to the Asiago Observatory. This procedure should not
have an impact on the parallax program but we search for correlations
with residuals in the solutions and these dates as a matter of
routine.

For each target we make one 30 second exposure, then measure both the
position and maximum pixel count for the target. The telescope is then
repointed so that the target is in the center of the CCD and the
exposure time is calculated in a simple manner to optimize the counts
for the target and reference stars. In this way we insure that spatial
variations on the CCD have minimal effect and that nightly variations
of the sky are to a reasonable extent accounted for. If the target is
not too bright or faint and the reference stars not too sparse, then
we choose a maximum pixel count that on a bad night will not require
more than a five minute exposure and on a good night defaults to this
optimal exposure time. We attempt to restrict all observations to
within 30 minutes of the meridian and to make two exposures for { each}
target per night.

\section{Calibrations}
The CCD images are processed using standard IRAF procedures. First all
frames are { bias} subtracted using the overscan regions and then a
flat is made for each night from a median of at least three sky
flats. For each object image we flag pixels within 95\% of the well
limit and these are not used in the calculation of the centroids. Each
image is divided by the median flat and then passed to an in-house
program that automatically produces an object list using a simple
three sigma above background criteria.

This object list and cleaned image is then processed to find the most
important measured quantity in this project - the centroid. In Smart
et al. (1999) \nocite{SMA99A} we discuss the experiments carried out
to determine the image centroids and conclude there that the best
procedure is an in-house two-dimensional gaussian profile fitter.  We
estimate by comparison of consecutive frames that the centroiding
precision is better than 0.02 pixels ($\sim$10 mas) for stars within 5
magnitudes of saturation and decreases rapidly for fainter stars.

The largest source of systematic error calibrated in MON92 was the
effect of Differential Color Refraction (DCR).  This occurs because
the frame defined by the reference stars has a different average color
than the target star. The position of the target star will be
refracted differently than the mean field by the atmosphere and the
size of this difference will be a function of the hour angle.  This is
especially important when observations were carried out far from the
meridian to maximize the parallax factor.

\begin{figure}[htp]
\epsfysize=6.0cm
\epsffile{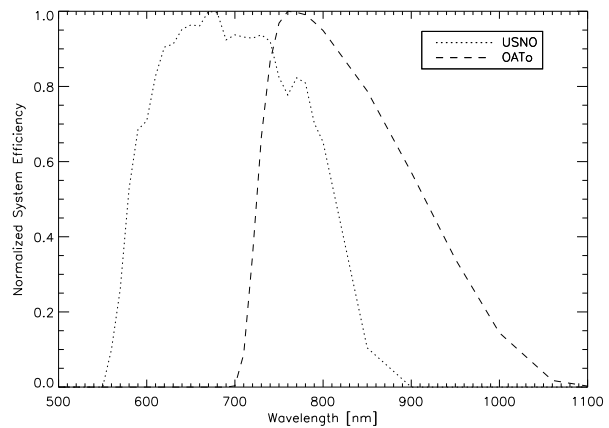}
\caption{The combined CCD quantum efficiency, optical transmission and
filter transmission for the USNO Flagstaff 1.5m reflector and the 
OATo 1.05m reflector.}
\label{calcbbandefflam_2.ps}
\end{figure}

In MON92 this effect was discussed at length and we are currently
using data taken over the course of the program to calibrate the
effect in a similar way. However we note that the combination of our
CCD and longer wavelength filter means that our DCR will be much lower
than theirs. In Fig. \ref{calcbbandefflam_2.ps} we plot the combined
CCD/filter/telescope efficency. In Smart et al. (1999) we show that
the DCR of the Torino program is 
{ negligible in the case of QSOs which have similar colors to these
white dwarfs and are close to the mean color of the reference
stars. As a result of our low DCR, the restriction that 
observations be made close to the meridian, and a low color difference
between the target and the mean reference star color, we believe we 
are justified in ignoring the } DCR correction for these particular
objects. Future work will include the DCR.

\section{Reduction Techniques}

\subsection{Determination of Relative Parallax}
For the relative parallaxes found { in this work} we have used a simple iterative
approach.  We will eventually use a more rigorous procedure that will
find directly the absolute parallax using the global central overlap
procedure similar to that described in Eichhorn
(1997)\nocite{EIC97A}. Since the stars under discussion here have
parallaxes of tens of milliarcseconds we expect refinements such
as the use of the central overlap to have minimal effect.

The selection of frames and reference stars is completely
automatic. We have the ability to manually select the reference stars
but testing on simulations has shown that an intelligent choice of
selection criteria is sufficient and has the advantage of being free of observer
bias. Initially all frames with an hour angle greater than one
hour and all frames used to set up the target are dropped.  A frame
from the middle of the sequence with a high number of stars is chosen
as the base-frame.

From the measured coordinates ($x,y$) of objects on the base frame we
calculate standard coordinates ($\xi,\eta$) by a linear transformation
using stars common to a reduction of the Digitized Sky Survey
(DSS). This sets the plate scale and orientation { which allows us
to use parallax factors in standard coordinates. The precision of this
} transformation is limited by the quality of the DSS solution; for
the orientation any errors less than 1 degree do not affect our use of
the standard coordinate parallax factors and we assume any errors in
the scale are less than 1\%.

Using pattern recognition software we find 6 stars in common with the
base frame and the other frames; any frame without at least six common
stars is dropped. We then orient each frame to the standard coordinate
system of the base frame 
using these 6 stars.  Then we find all stars in common with the base
frame and pick the 12 most frequent stars including the target star,
all frames without these stars are dropped. These 12 stars define the
reference system for the target star. Obviously if all the frames have
a larger subset of stars in common then we will use that, but usually
to have at least 12 stars we have to drop some of the frames.

We then find the frame parameters $A, B, C$ using the common stars by
solving the transformation:
\begin{equation}
 \xi_{\mu,\nu_{o}} = A_{\nu} x_{\mu,\nu} + B_{\nu} y_{\mu,\nu} +C_{\nu}
\label{xiplate} 
\end{equation} 
where $\mu,\nu$ represent the star and frame number respectively and
the $\nu_o$ indicates the base plate. Here, and in further discussion,
we develop the procedure in the $\xi$ coordinate but the relations are
analogous in $\eta$.  In calculating the parameters in Eq. (1) the
target star is not used.

Then for all stars on the base frame we find a preliminary parallax
and proper motion using:
\begin{equation}
  \xi_{\mu,\nu} = \xi_{\mu,\nu_{o}}(1.0+a_{\mu}) +
	(t_{\nu}-t_{\nu_{o}}) \mu_{\mu} +
	(P_{\xi_{\nu}}-P_{\xi_{\nu_{o}}})\omega_{\mu}
\label{xiseq} 
\end{equation}
Where: $a_{\mu}$ allows for the error in the base plate position,
$(t_{\nu}-t_{\nu_{o}}) $ is the time difference,
$(P_{\xi_{\nu}}-P_{\xi_{\nu_{o}}}) $ is the parallax factor difference
and $\xi_{\mu,\nu_{o}}, \mu_{\mu}$ and $\omega_{\mu}$ are the
position, proper motion and parallax of the star ${\mu}$.  The
parallax factors are given by: $ P_{\xi} =
\cos\alpha~\sin\lambda~\cos\epsilon - \sin\alpha~\cos\lambda $ and $
P_{\eta} = (\sin\epsilon~\cos\delta - \cos\epsilon~\sin\alpha
~\sin\delta)\sin\lambda - \cos\alpha~\sin\delta~\cos\lambda $ where
$\alpha$, $\delta$ are the fields equatorial coordinates, $\lambda$ is
the solar longitude, and $\epsilon$ the inclination of the ecliptic.

From this fit we have predicted positions of the reference stars which
we can then use to iterate again through Eqs. (\ref{xiplate}) and
(\ref{xiseq}). Before doing this iteration we attempt to identify bad
frames and stars by finding the root mean square of the residuals per
frame and per star. Every frame or star with a root mean square
greater than three sigma from the mean we delete from the next
iteration. 

We iterate for at least 4 iterations and then stop when
the change in the parallax is less than one mas. We
attempted various forms of reweighting as we iterated both in the
calculation of the plate parameters and in the calculation of the
final parallax - however we found the best results by only applying
weights in the calculation of the plate parameters in the final
iteration.  We have also reduced the sequences using Gaussfit
(\citealt{jef88}), constraining the parallaxes and proper motions of
the reference stars to zero, and this provide similar results,
increasing our confidence in the solutions.

Using this reduction procedure we have implicitly assumed that the
proper motions and parallaxes of the reference stars are uncorrelated
within the frame. This will not be true for the proper motions, which
in a given direction will be dominated by a combination of
differential galactic rotation and solar motion, but this will only
add a systematic error to the proper motions and we can ignore it as
our main goal here is the parallax.

\subsection{Correction to an absolute parallax}
The quantity calculated above is the parallax of the target star
relative to the anonymous reference stars. For each reference star we
have also calculated a parallax and proper motion using the same
reference system as that of the parallax star. Inevitably the
reference star parameters ($a_{\mu}, \mu_{\mu}$, $\omega_{\mu}$) are
correlated with the definition of the reference system; however with at
least 12 stars in each system we consider this correlation to be
negligible.

{ From} the target star's relative parallax we should subtract the
mean parallax of the reference stars weighted by the weights used in the plate
solutions and the dependences (\citealt{Sch11}).  This enforces the
assumption that the reference frame is at a mean parallax of zero and
the use of weights in this step allows us to consider
each reference star as having unit weight in the final step for the
correction from relative to absolute parallax. However, the dependences
are all nearly equal and  we expect this mean to be zero. We have found that
usually the mean is much smaller than the variance of the various distances
and we have decided not to apply this correction as it is not
sensibly different from zero.

{ 
As part of the calibration we observed standard magnitudes for all
stars in each field. Using the Mendez \& van Altena
(1996\nocite{men96}) galaxy model we can calculate the most probable
distance of each reference star based on their magnitude. The mean of
these distances is an estimate of the correction to absolute
parallax that we must add to the calculated targets relative parallax.
In Table 1 we list the range and mean of the reference star magnitudes
and the corresponding relative to absolute correction (RAC) for each
field.
}
\begin{table}[h]
\centering
\caption{Magnitude ranges and means in the I filter, and
relative to absolute corrections in milliarcseconds as found from the
Mendez \& van Altena galaxy model.}
\leavevmode
\begin{tabular}{ccccc}
\hline  \\
Field   & $I_{min}$ & $I_{max}$ & $<I_{w}>$ & RAC \\
\hline  \\
0322$-$019 & 11.21 &  15.88 & 14.37  & 2.0  \\ 
0423$+$120 & 12.05 &  15.81 & 14.69  & 1.3  \\ 
1126$+$185 & 11.80 &  15.90 & 14.49  & 2.1  \\ 
1716$+$020 & 10.97 &  14.76 & 12.95  & 1.2  \\ 
1840$-$111 & 9.650 &  14.43 & 12.88  & 0.4  \\ 
2216$+$484 & 10.39 &  15.48 & 14.13  & 0.8  \\ 
\hline \\
\end{tabular}
\end{table} 

\subsection{Determination of errors}
After the final iteration the formal error of the star parameters come
directly from the final covariance matrix. As we used a nominal input
for the centroiding error we multiply the terms of the covariance
matrix by the error of unit weight.

Based on various considerations (see for example Smart et al, 1997)
\nocite{SMA97D} we estimate the error in the RAC found using a galaxy
model to be about 30\%. Therefore to the formal relative parallax
error above we add in quadrature 30\% of the RAC to provide us with a
final error.

\begin{table*}[ht]
\centering
\caption{Parallaxes and proper motions for the six white dwarf
candidates in the Torino Program. See text for explanation of
columns. }
\leavevmode
\begin{tabular}{crrrrrrrrrr}
\hline  \\
WD Name & $N_s$ & $N_f$ & $\omega $ mas ~~~~~& $\mu_{\alpha} $  mas/yr ~~
&  $\mu_{\delta} $ mas/yr ~~& V~~~~~~ & V-I~~~~ & RMS mas & $\Delta$T years \\
\hline  \\
0322$-$019 &  13 &  83 &   59.5$\pm$3.2 &  222.5$\pm$1.0 & -857.5$\pm$1.2 & 16.11$\pm$0.06  &  0.87$\pm$0.11 & 18.07  &  6.20 \\ 
0423$+$120 &  24 &  56 &   57.6$\pm$2.5 & -106.7$\pm$1.0 & -225.5$\pm$0.9 & 15.43$\pm$0.04  &  0.62$\pm$0.05 & 10.94  &  5.31 \\ 
1126$+$185 &  11 &  41 &    0.9$\pm$2.4 &    6.0$\pm$1.0 &    7.2$\pm$1.1 & 13.97$\pm$0.04  &  0.75$\pm$0.05 &  9.48  &  6.09 \\ 
1716$+$020 &  21 & 159 &   19.0$\pm$2.3 & -439.8$\pm$0.7 & -279.2$\pm$0.9 & 14.30$\pm$0.10  & -0.02$\pm$0.17 & 18.49  &  6.27 \\ 
1840$-$111 &  77 &  64 &   34.9$\pm$5.1 & -234.7$\pm$1.6 & -247.7$\pm$1.5 & 14.18$\pm$0.09  &  0.38$\pm$0.14 & 21.07  &  6.15 \\ 
2216$+$484 & 130 &  90 &   33.6$\pm$2.5 &  146.7$\pm$1.2 &  -40.3$\pm$0.8 & 16.13$\pm$0.10 &  0.85$\pm$0.15 & 15.19  &  5.04 \\ 
\hline \\
\label{results}
\end{tabular}
\end{table*} 

We note that the use of the DSS to transfer the frame to the sky is
basically a setting of the scale and orientation of the CCD. From an
examination of the plate parameter errors and tests against external
catalogs we have found that the scale of the DSS is 
correct to better than 1\%. This error is invisible to the calculation
of the relative parallax and cannot be easily estimated on a case by
case basis. We consider it a systematic error that for all of the
targets here will be below the value of the formal error (being at
most 1\% of the targets parallax) and therefore if the user requires
a maximum error estimate we suggest that they add 1\% of the targets
parallax to the quoted error.

The error on the proper motions is more difficult to access as it is 
affected directly by the orientation, scale, and the fact that
they are relative as we have made no attempt to account for systematic
motions of the reference stars. Any use of the proper motions quoted
here that requires high precision should be done with care.
 
\section{Results}
In the TOPP program there are six white dwarf candidates for which we
present the results here. These have been chosen because they will not
be affected by the DCR which is currently not calibrated in the
reduction procedure. In Table \ref{results} we list the objects,
number of reference stars $N_s$, number of observations used in the
final fit $N_f$, absolute parallaxes $\omega $, proper motions
$\mu_{\alpha}, \mu_{\delta}$, magnitude V, color V-I,{ ~the root mean 
square of the residuals between the observations and the final fit 
RMS} and the duration of the
observational sequence in years $\Delta$T.

\begin{figure}[htp]
\epsfysize=8.0cm
\epsffile{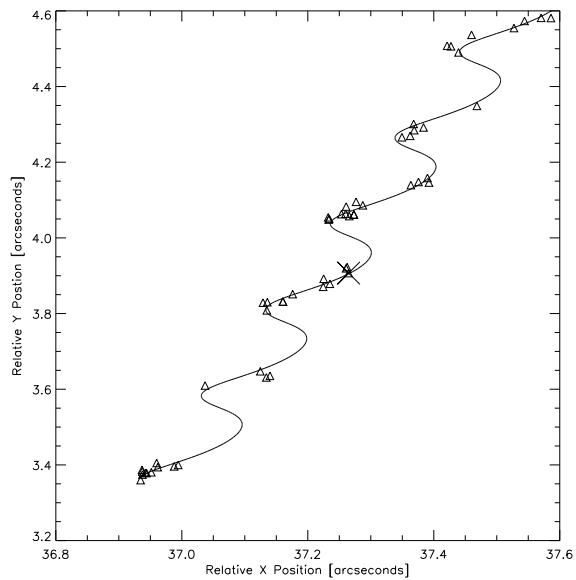}
\caption{Observations of 0423+120 along with the fitted path. The X
marks the position of the base plate.}
\label{n1411_par}
\end{figure}

In Fig. \ref{n1411_par} we present as a typical example the observed
and fitted path for the target 0423+120. The average residual is 12
and 10 milliarcseconds in $\xi$ and $\eta$ respectively.

Two targets, 1716+020 and 1840$-$111, are part of wide binary systems
with M-dwarf stars. In both cases we find the difference in proper
motions measured in these pairs are less than the errors. As can be
seen from Table \ref{doublerror} both these systems have parallaxes in
good agreement.  The systematic direction of the existing difference,
where the red star has larger parallax, is consistent with what is
expected by not applying the DCR. Nevertheless the difference is of
the order of the errors justifying our neglect of the DCR for these
objects.
\begin{table}[ht]
\centering
\caption{Parallaxes of two binary systems.}
\leavevmode
\begin{tabular}{rrrr}
\hline  \\
Target  & $\omega $ mas~~~~ & V & V-I \\
\hline  \\
1840$-$111  & 34.9$\pm$5.1 & 14.18  &  0.38 \\ 
Companion   & 37.9$\pm$4.0 & 15.14  &  3.23 \\
1716$+$020  & 19.0$\pm$2.3 & 14.30  & -0.02 \\ 
Companion   & 21.6$\pm$1.3 & 14.28  &  2.93 \\ 
\hline \\
\label{doublerror}
\end{tabular}
\end{table}

\subsection{Comments on individual targets}

{\bf 0322$-$019} (LHS1547) is listed in the Astronomisches Rechen-Institut
database for nearby stars
(http://www.ari.uni-heidelberg.de/aricns/index.htm, hereafter ARICNS)
as a DZ9 with a photometric parallax of 62 mas, consistent
with our measurement of 59.5$\pm$3.2.

{\bf 0423+120} (EJ 169) is listed in the ARICNS as a DC8 with a photometric
parallax of 81$\pm$11 mas, significantly different from our value of
58 mas. This is derived from the absolute magnitude of McCook \& Sion
(1999\nocite{mcc99}, hereafter MCC99) and the apparent magnitude in the
ARICNS.

{\bf 1126+185} (GJ 3667) is listed as a DC9 with a photometric
parallax of 114$\pm$13 mas, inconsistent with our zero value.  
{ We have used the finding chart provided in
Green et al (1986\nocite{gre86}) where it was first identified }
as a composite DC white dwarf based
on at least two spectra which were not published. Subsequent work with
new spectroscopy by Putney (1997\nocite{put97}, Fig. 2l) also
confirmed this result with a spectra that indicated ``a composite of a
DC and a cool star''. We have discussed these observations with
various spectroscopists in light of this object's near zero proper
motion and parallax, but have not been able to come up with other
suggestions. { As no other star in the TOPP field has a significant 
proper motion we therefore assume this is a case of
misclassification and/or misidentification.}

{\bf 1716+020} (YPC 3926.01A, LHS3278) is listed in the Fourth Yale
Parallax Catalog (\citealt{vana01}) with a mean parallax of 28$\pm$2
mas and a spectral type of DA6 which is inconsistent with
our value of 19$\pm$2 mas. However, as we noted before,
this object has a companion red dwarf (LHS3279) and to calculate the
Yale mean parallax the various observations of both stars were
averaged. In Table \ref{ypc} we list the individual observations.
\begin{table}[h]
\centering
\caption{Observations from the Fourth Yale Parallax Catalogue for the
determination of the parallax of 1716+020.}
\leavevmode
\begin{tabular}{ccc}
\hline  \\
Observatory  & 1716+020 & LHS3279  \\
\hline  \\
Lick  & 24.4$\pm$0.9 & 33.2$\pm$0.7 \\
Sproul  & 13.9$\pm$1.4 & 45.9$\pm$1.6 \\
USNO  & 24.2$\pm$0.4 & 31.4$\pm$0.4 \\
OATo & 19.0$\pm$2.3 & 21.6$\pm$1.3 \\
\hline \\
\label{ypc}
\end{tabular}
\end{table} 
Considering 1716+020 alone, our measurement is consistent with its
mean parallax, whereas inconsistency arises when the parallax of the
red dwarf companion is considered. This difference may be due to an
incorrect treatment of the refraction effect in the other programs
where it is much more important than ours.  We also note that the
projected distance in all cases is less than 0.016 parsecs while only
our values can produce distances that are consistent with a line of
sight difference of this order.

{\bf 1840-111} (GJ 2139) is listed in the ARICNS as a DA5 with a
photometric parallax of 53$\pm$6 mas, inconsistent with our value of
35mas. This is a derived value from the MCC99 work and is discussed
below. The higher error of this target may be a result of it's lower
declination combined with the high field density, which could adversely
affect the centroiding by the crowding when the seeing was not
optimal.

{\bf 2216+484} (GD 402) has no parallax estimates in the literature.
In Bergeron et al. (1990)\nocite{ber90}, based on anomalies in the
spectra, they conclude that this object is probably a composite system
of a DA and a DC white dwarf. From the apparent magnitude and our
parallax we derive an absolute magnitude of 13.76 in V which is
consistent with a single DA star and contrary to the thesis of a
double system which is estimated in Bergeron et al. to be of absolute
magnitude 12.94. An examination of the point spread function during
good seeing does not show any indication of duplicity. The apparent
magnitude and the astrometric residuals of the system are stable over
the program. These observations are not conclusive proof of
singularity but they do put restrictions on orbit orientation in the
case of a composite system.

\subsection{External Comparisons}
In Table \ref{mccook} we list the absolute magnitudes derived from the
parallaxes and apparent magnitudes quoted here along with those
derived in MCC99 from spectrophotometry. We have not included
1126$+$185 and we note that the 2216$+$484 MCC99 value is based on the
assumption that it is a single DA7 star.
\begin{table}[h]
\centering
\caption{Absolute magnitudes calculated from TOPP parallaxes and
apparent magnitudes along with those derived in MCC99 from
spectrophotometry, the code indicates the method used: 2=multichannel
spectrophotometric colors, 3 = uvby colors.}
\leavevmode
\begin{tabular}{cccc}
\hline  \\
Target     &  TOPP & MCC99 & code \\
\hline  \\           	   
0322$-$019 & $14.98^{ +.11}_{ -.12}$ & 15.10 & 2 \\
0423$+$120 & $14.23^{ +.09}_{ -.10}$ & 15.05 & 3 \\
1716$+$020 & $10.69^{ +.25}_{ -.28}$ & 11.78 & 3 \\
1840$-$111 & $11.90^{ +.30}_{ -.34}$ & 12.81 & 3 \\
2216$+$484 & $13.76^{ +.16}_{ -.17}$ & 13.57 & 2 \\
\hline \\
\label{mccook}
\end{tabular}
\end{table} 
It is beyond the scope of this paper to discuss the relative merits of
absolute magnitudes derived from spectrophotometry versus those from parallaxes
and the sample here is too small for such a comparison. However, there is a
suggestion that the absolute magnitudes derived from uvby colors are
underestimates { (i.e. numerically larger) than the true magnitudes - 
assuming that there is not a systematic error of 30\% in our results.}

\begin{figure}[htp]
\epsfysize=8.0cm
\epsffile{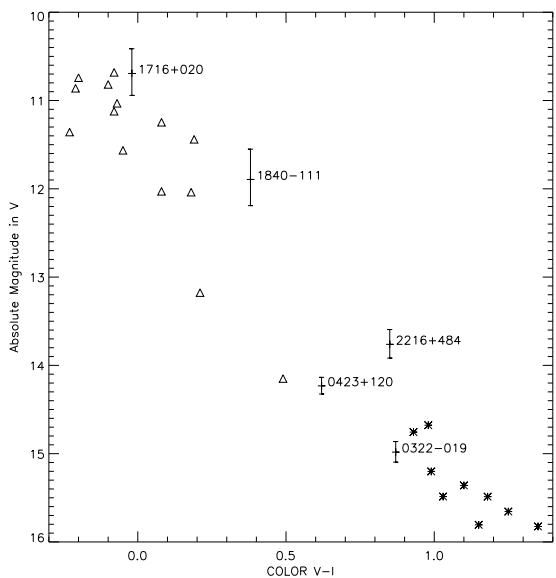}
\caption{The white dwarf sequence in the HR Diagram. The asterisks are from
MON92, triangles from HIPPARCOS and pluses with error bars this paper.}
\label{all_surveys5}
\end{figure}

Apart from 1716+020 we do not have any direct comparisons to other
trigonometric parallaxes determinations.  We can however make an
indirect comparison.  In Fig. ~\ref{all_surveys5} we plot the white
dwarf tract in the H-R Diagram as derived from both HIPPARCOS stars
and those published in MON92 with the TOPP results over-plotted. We
believe this adequately shows that our results are consistent with
other high precision parallax determinations.

\section{Future Work}
As we have stated the reduction technique used here is not
optimal. The central overlap method developed by Eichhorn (1997) and
extended by Rapaport (1998\nocite{rap98}) is preferred. This method
will allow the simultaneous adjustment of frame parameters, star
parameters and the correction to the absolute frame using all the
information available. It will mean that we will not have to use
formal errors as weights, assume a mean parallax for the reference
frame or neglect the dependencies. Significant external information is
required for this procedure and it will probably
only be important for stars which have parallaxes much closer to our
precision limit where the relative error is larger. We do plan to
implement this procedure in the future development of this program.

\section{Conclusion}
We have shown that the Torino program is able to produce useful
parallaxes with precisions and in a magnitude range that complement
the HIPPARCOS results and the USNO program. We plan to improve the
reduction procedure which will only improve the precision and produce
parallaxes for the other 100+ targets observed over the last 6 years.
Once the automatization of the REOSC telescope is complete we expect to
be able to continue the program looking at the many exotic objects
being discovered by the large all-sky surveys in progress.

We have measured the first parallax for five of six white dwarfs
candidates. One object, 1126$+$185, is obviously misclassified,
another, 2216$+$484, is inconsistent with previous observations
that suggest it is a composite system.  The overall photometric
magnitude calibration of McCook \& Sion (1999\nocite{mcc99}) appears
to overestimate the objects magnitude. These discrepancies with
published work highlight the need for more direct parallax
determinations which are fundamental tests of current theories.

\section{Acknowledgments}

First thanks go to the Consiglio Nazionale delle Ricerche for initially
providing the CCD, and to the former Consiglio per la Ricerca
Astronomica current INAF and the Directors of OATo for
their longterm support of this program.  A six year observational program could
not have been carried out without the support of a number of the
OATo staff. In particular we wish to thank Massimo Villata,
Francesco Porcu, Fillipo Racioppi, Leonardo Corcione, Marco Del B\`o,
Francesco Salvati and the QSO monitoring group for observations made
in service mode. Thanks to Bill Jefferys and Michel Rapaport
for many useful discussions, also Dave Monet and Conrad Dahn for
providing us with an initial kick start for this project. { We would
also like to thank the referee, Dr van Leeuwen, whose comments
helped improve the clarity of this paper.} Finally, RLS
is grateful to Heinrich Eichhorn who before his death was a constant
source of guidance and through his work continues to inspire
today. RLS acknowledges the past support of the Royal Society and
Torino University.


\begin{thebibliography}{0}
\expandafter\ifx\csname natexlab\endcsname\relax\def\natexlab#1{#1}\fi

\end{thebibliography}


\begin{thebibliography}{15}
\expandafter\ifx\csname natexlab\endcsname\relax\def\natexlab#1{#1}\fi

\bibitem[{{Bergeron} {et~al.}(1990){Bergeron}, {Liebert}, \&
  {Greenstein}}]{ber90}
{Bergeron}, P., {Liebert}, J., \& {Greenstein}, J.~L. 1990, Astrophys. J., 361,
  190

\bibitem[{{Dahn}(1998)}]{DAH98}
{Dahn}, C.~C. 1998, in IAU Symp. 189: Fundamental Stellar Properties, Vol. 189,
  19

\bibitem[{{Eichhorn}(1997)}]{EIC97A}
{Eichhorn}, H. 1997, Astron. Astrophys., 327, 404

\bibitem[{Gatewood(1987)}]{GAT87}
Gatewood, G. 1987, Astron. J., 94, 213

\bibitem[{{Green} {et~al.}(1986){Green}, {Schmidt}, \& {Liebert}}]{gre86}
{Green}, R.~F., {Schmidt}, M., \& {Liebert}, J. 1986, Astrophys. J., Suppl.
  Ser., 61, 305

\bibitem[{{Jefferys} {et~al.}(1988){Jefferys}, {Fitzpatrick}, \&
  {McArthur}}]{jef88}
{Jefferys}, W.~H., {Fitzpatrick}, M.~J., \& {McArthur}, B.~E. 1988, Celestial
  Mechanics, 41, 39

\bibitem[{{McCook} \& {Sion}(1999)}]{mcc99}
{McCook}, G.~P. \& {Sion}, E.~M. 1999, Astrophys. J., Suppl. Ser., 121, 1

\bibitem[{{Mendez} \& {van Altena}(1996)}]{men96}
{Mendez}, R.~A. \& {van Altena}, W.~F. 1996, Astron. J., 112, 655

\bibitem[{Monet {et~al.}(1992)Monet, Dahn, Vrba, Harris, Pier, Luginbuhl, \&
  Ables}]{MON92A}
Monet, D.~G., Dahn, C.~C., Vrba, F.~J., {et~al.} 1992, Astron. J., 103, 638

\bibitem[{{Putney}(1997)}]{put97}
{Putney}, A. 1997, Astrophys. J., Suppl. Ser., 112, 527

\bibitem[{{Rapaport}(1998)}]{rap98}
{Rapaport}, M. 1998, Astron. Astrophys., 335, 769

\bibitem[{{Schlesinger}(1911)}]{Sch11}
{Schlesinger}, F. 1911, Astrophys. J., 33, 161

\bibitem[{{Smart} {et~al.}(1999){Smart}, {Bucciarelli}, {Lattanzi}, {Massone},
  \& {Chiumiento}}]{SMA99A}
{Smart}, R.~L., {Bucciarelli}, B., {Lattanzi}, M.~G., {Massone}, G., \&
  {Chiumiento}, G. 1999, Astron. Astrophys., 348, 653

\bibitem[{Smart {et~al.}(1997)Smart, Lattanzi, \& Drimmel}]{SMA97D}
Smart, R.~L., Lattanzi, M.~G., \& Drimmel, R. 1997, in Wide-field Spectroscopy,
  ed. E.~e.~a. Kontizas (Kulwer Academic Press), 195

\bibitem[{{van Altena} {et~al.}(2001){van Altena}, {Lee}, \&
  {Hoffleit}}]{vana01}
{van Altena}, W.~F., {Lee}, J.~T., \& {Hoffleit}, E.~D. 2001, {The General
  Catalogue of Trigonometric Stellar Parallaxes, Fourth Edition} (L. Davis
  Press)

\end{thebibliography}
\bibliographystyle{../../aa/bibtex/aa}

\end{document}